\documentclass[11pt]{article}

\PassOptionsToPackage{table}{xcolor}
\usepackage[preprint]{acl}

\usepackage{times}
\usepackage{latexsym}
\usepackage[T1]{fontenc}
\usepackage[utf8]{inputenc}
\usepackage{microtype}
\usepackage{inconsolata}
\usepackage{graphicx}
\usepackage{amsmath,amssymb,mathtools}
\usepackage{booktabs,multirow,tabularx,array}
\usepackage{placeins}
\usepackage{enumitem}
\usepackage{algorithm}
\usepackage{algorithmic}
\usepackage{tikz}
\usetikzlibrary{arrows.meta,backgrounds,calc,fit,positioning,shapes.geometric,shapes.multipart}
\usepackage{pgfplots}
\pgfplotsset{compat=1.18}
\usepackage[capitalize,noabbrev]{cleveref}

\usepackage{amsmath,amsfonts,bm}

\def\eqref#1{equation~\ref{#1}}
\def\1{\bm{1}}

\DeclareMathAlphabet{\mathsfit}{\encodingdefault}{\sfdefault}{m}{sl}
\SetMathAlphabet{\mathsfit}{bold}{\encodingdefault}{\sfdefault}{bx}{n}

\definecolor{verasblue}{HTML}{2F6B9A}
\definecolor{verasteal}{HTML}{2A9D8F}
\definecolor{verasgold}{HTML}{E9A23B}
\definecolor{verasred}{HTML}{C94C4C}
\definecolor{verasgray}{HTML}{5F6B76}
\definecolor{veraseasybg}{HTML}{EAF5EC}
\definecolor{verasmediumbg}{HTML}{FFF4D8}
\definecolor{verashardbg}{HTML}{FBE9E7}
\definecolor{placeholderbg}{HTML}{FFF5D6}

\newcommand{\method}{\textsc{Veras}}

\title{VERAS: Verification-Guided Adaptive Search for Reliable Operations Research}
\author{
    \textbf{Minbo Ma}\textsuperscript{1}
    \thanks{Equal contribution.},
    \textbf{Honglin Zhang}\textsuperscript{1}
    \footnotemark[1],
    \textbf{Bangtao Zhou\textsuperscript{2}},
    \textbf{Chunguang Yu\textsuperscript{2}},
    \\
    \textbf{Dasheng Ju\textsuperscript{2}},
    \textbf{Yichen Shen\textsuperscript{3}},
    \textbf{Fengli Xu\textsuperscript{1}},
    \textbf{Yong Li\textsuperscript{1}}
    \thanks{Corresponding author:liyong07@tsinghua.edu.cn}
    \\[0.6em]
    \textsuperscript{1}Tsinghua University, Beijing, China \\
    \textsuperscript{2}Meituan, Beijing, China \\
    \textsuperscript{3}Shanghai Jiaotong University, Shanghai, China
}

\begin{document}
\maketitle

\begin{abstract}
Advances in reasoning, coding, and tool use enable LLM agents to solve operations research (OR) problems through iterative problem modeling, program construction, and verification. 
Yet passing self-authored verification does not ensure that the solutions satisfy task constraints and meet optimization quality.
This gap can arise from \emph{correlated specification errors}, where the generated program and its checks share omissions or misinterpretations of task requirements, alongside incomplete verification coverage and ineffective optimization strategies.
We propose VERAS, a \textbf{ver}ification-guided \textbf{a}daptive \textbf{s}earch harness for reliable OR that uses verification evidence to prioritize feasibility repair, constraint-coverage review, and solution-quality improvement.
VERAS combines task-specific solution checkers with a separate verification agent to examine candidate solutions against the task requirements. It uses verification feedback to guide feasibility repair, additional constraint checking, and objective improvement, while candidate acceptance criteria screen revisions for deterioration in output validity or constraint satisfaction. 
On ORAgentBench, VERAS achieves a 42.99\% pass rate, exceeding the reported 35.51\% of Codex using the same LLM. Across two industrial-scale OR problems, VERAS reduces inter-datacenter peak bandwidth by 10.34\% relative to native Codex and lowers annualized distribution network planning costs by 6.28\% compared with expert-designed solutions.
Code and evidence are available at \url{https://github.com/holimario/VERAS}.
\end{abstract}

\section{Introduction}
\label{sec:introduction}
Operations research (OR) underpins decision optimization across diverse domains, such as transportation dispatch~\citep{vazifeh2018addressing}, energy planning~\citep{an2025repositioning}, and computing workload scheduling~\citep{colangelo2026ai}.
Realizing these benefits requires an iterative workflow spanning task understanding, problem formulation, model solving, solution validation, and model refinement~\citep{williams2013model}.
This iterative process is expertise-intensive, requiring substantial domain knowledge and optimization expertise at each stage.

Recent advances in large language models (LLMs) reasoning, coding, and tool use have opened new opportunities for automated OR problem solving~\citep{yao2022react,guo2025deepseek}.
Early studies focus on translating problem descriptions into mathematical formulations and solver code, supported by domain-specific training~\citep{huang2024orlm}. 
Agentic systems such as CoE~\citep{xiao2024chain}, OptiMUS~\citep{ahmaditeshnizi2024optimus}, and NEMO~\citep{song2026nemo} have extended these capabilities toward end-to-end OR solving by integrating problem modeling, program execution, and feedback-driven refinement. 
In parallel, FunSearch~\citep{romeraparedes2024funsearch} and AlphaEvolve~\citep{novikov2025alphaevolve} leverage evaluation-guided program evolution to discover and improve heuristics and algorithms.
While these approaches differ in their emphasis on mathematical modeling and solver-based execution versus algorithm design, both rely on executable programs and iterative evaluation feedback.

\begin{figure*}[th]
\centering
\includegraphics[width=0.8\linewidth]{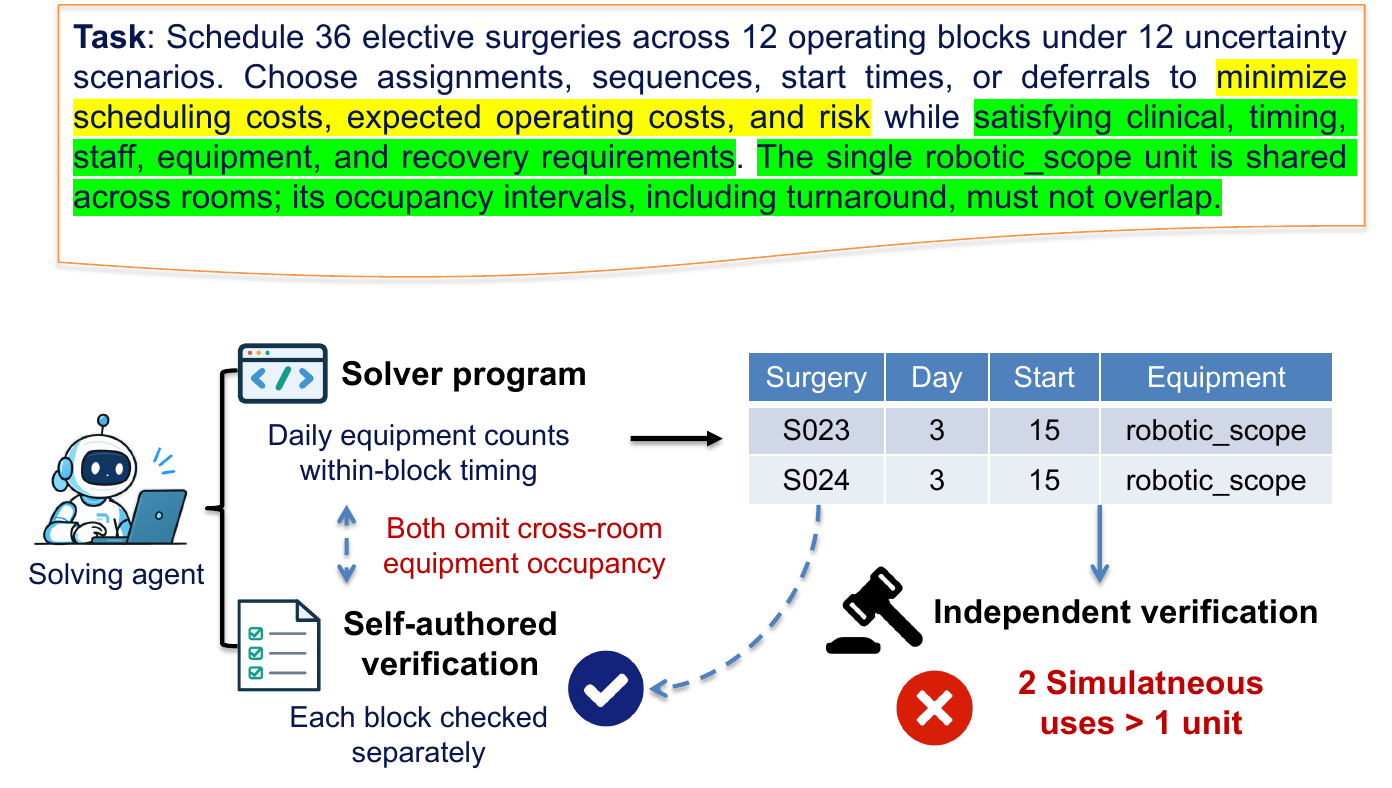}
\caption{Correlated specification errors in OR agent. The optimization objective (yellow), the task constraints (green). The solver program and its self-authored verification both omit cross-room equipment occupancy, allowing an infeasible schedule to pass internal checks but failing in independent evaluation.}
\label{fig:problem}
\end{figure*}

The effectiveness of such refinement depends on the reliability of evaluation feedback. 
When OR agents build their own checks to verify candidate solutions, these self-authored verifications may inherit the interpretation of task requirements as the generated program. 
This creates the risk of \emph{correlated specification errors}, where the program and its checks share the same omissions or misinterpretations, allowing incorrect solutions to pass internal verification. Figure~\ref{fig:problem} illustrates this failure through an agent trajectory on ORAgentBench~\citep{li2026oragentbench}, in which both the generated optimization program and its self-authored checks overlook shared-equipment occupancy across operating rooms, allowing the same equipment unit to be assigned to overlapping surgeries. 
Such feedback hardly exposes the shared omission and therefore provides no basis for correcting it. 
This raises a broader question: \emph{how can verification grounded in task requirements guide agents to correct errors and improve solution quality?}

Addressing this question involves three challenges. First, \textbf{self-authored verification is susceptible to incomplete coverage}. Generated programs and verification may share specification errors, leaving some task requirements unexamined and internal verification results inconclusive. Second, \textbf{verification feedback often provides insufficient guidance for refinement}. Limited diagnostic evidence makes it difficult to distinguish modeling errors, unverified requirements, and opportunities for solution-quality improvement. Third, \textbf{iterative revisions may undermine previously verified properties}. Revisions intended to repair errors or improve the objective may introduce new violations or degrade solution quality, undermining previously established progress.

To address these challenges, we propose \textbf{VERAS}, a verification-guided adaptive search harness for reliable OR problem solving. Built upon a general-purpose coding agent, VERAS checks candidate solutions against task requirements using output validity checks and task-specific solution checkers, and conditionally invokes a separate verification agent to examine requirements beyond existing checks. Candidate status and verification feedback guide feasibility repair and solution-quality improvement through revisions to models, algorithms, and programs. Candidate acceptance criteria then compare output validity and checked constraint violations with the previous candidate, retaining the update or restoring the previous solution. These mechanisms integrate solution verification, targeted revision, and candidate selection into a unified solving workflow.

Our contributions are threefold:
\begin{itemize}[leftmargin=1.25em]
    \item We identify correlated specification errors in program construction and self-authored verification, and introduce task-grounded verification to expose constraint violations and coverage gaps.
    \item We introduce VERAS, a verification-guided adaptive search harness that uses verification evidence to steer OR solving refinement while rejecting updates that worsen verified constraint satisfaction.
    \item We evaluate VERAS on ORAgentBench and two industrial OR case studies spanning heuristic algorithm design and mathematical programming, demonstrating improvements in task success and solution quality.
\end{itemize}

\section{Related Work}
\paragraph{LLMs for OR Modeling and Algorithm Design.}
Recent LLM-based approaches to operations research (OR) have explored different ways of constructing executable optimization programs. In mathematical programming, CoE~\citep{xiao2024chain} coordinates specialized agents for modeling and code generation, while OptiMUS~\citep{ahmaditeshnizi2024optimus} and OR-LLM-Agent~\citep{zhang2025orllmagent} integrate formulation, solver execution, and iterative refinement. Another line of research focuses on algorithm design, where EoH and ReEvo employ LLM-guided evolution to develop optimization heuristics~\citep{liu2024eoh,ye2024reevo}. Despite their different approaches, both lines of work construct executable programs that produce optimization decisions and can be iteratively improved through feedback.

\paragraph{Evaluation-Guided Program Search and Verification.}
Evaluation feedback plays a central role in refining generated programs. FunSearch and AlphaEvolve employ automated evaluators to guide evolutionary search toward improved algorithms~\citep{romeraparedes2024funsearch,novikov2025alphaevolve}. In automated OR modeling, verification additionally concerns whether generated formulations and decisions faithfully satisfy task requirements. NEMO employs independently generated optimization and simulation programs for execution-based validation~\citep{song2026nemo}, while TriVAL and Opt-Verifier examine modeling correctness through stage-wise validation and complementary checks of model structure and solutions~\citep{fang2026trival,liu2026optiver}. ORAgentBench further assesses end-to-end OR agents through independent evaluation of decision feasibility and quality~\citep{li2026oragentbench}. These studies establish the importance of execution feedback and verification, while \method{} focuses on how verification evidence can guide subsequent search. By distinguishing witnessed violations from gaps in verification coverage, \method{} adapts program refinement and controls candidate acceptance based on the available evidence.
\section{VERAS}
\label{sec:method}

\subsection{Overview}
\label{sec:overview}
VERAS guides iterative OR problem solving by distinguishing detected constraint violations from incomplete verification coverage, thereby directing subsequent actions toward feasibility restoration, additional auditing, or objective improvement.

Given an OR instance $\mathcal I$ comprising a task description and instance data, we formulate the optimization problem as
\begin{equation}
\begin{aligned}
\min_{\mathbf{x}} \quad
& f(\mathbf{x};\mathcal I)\\
\mathrm{s.t.}\quad
& c_j(\mathbf{x};\mathcal I)=1,
\quad j\in\mathcal J,
\end{aligned}
\label{eq:or_problem}
\end{equation}
where $\mathbf{x}$ denotes the decision variables, $f$ the objective, and $c_j$ the constraint satisfaction predicates.

\begin{figure*}[t]
\centering
\includegraphics[width=\linewidth]{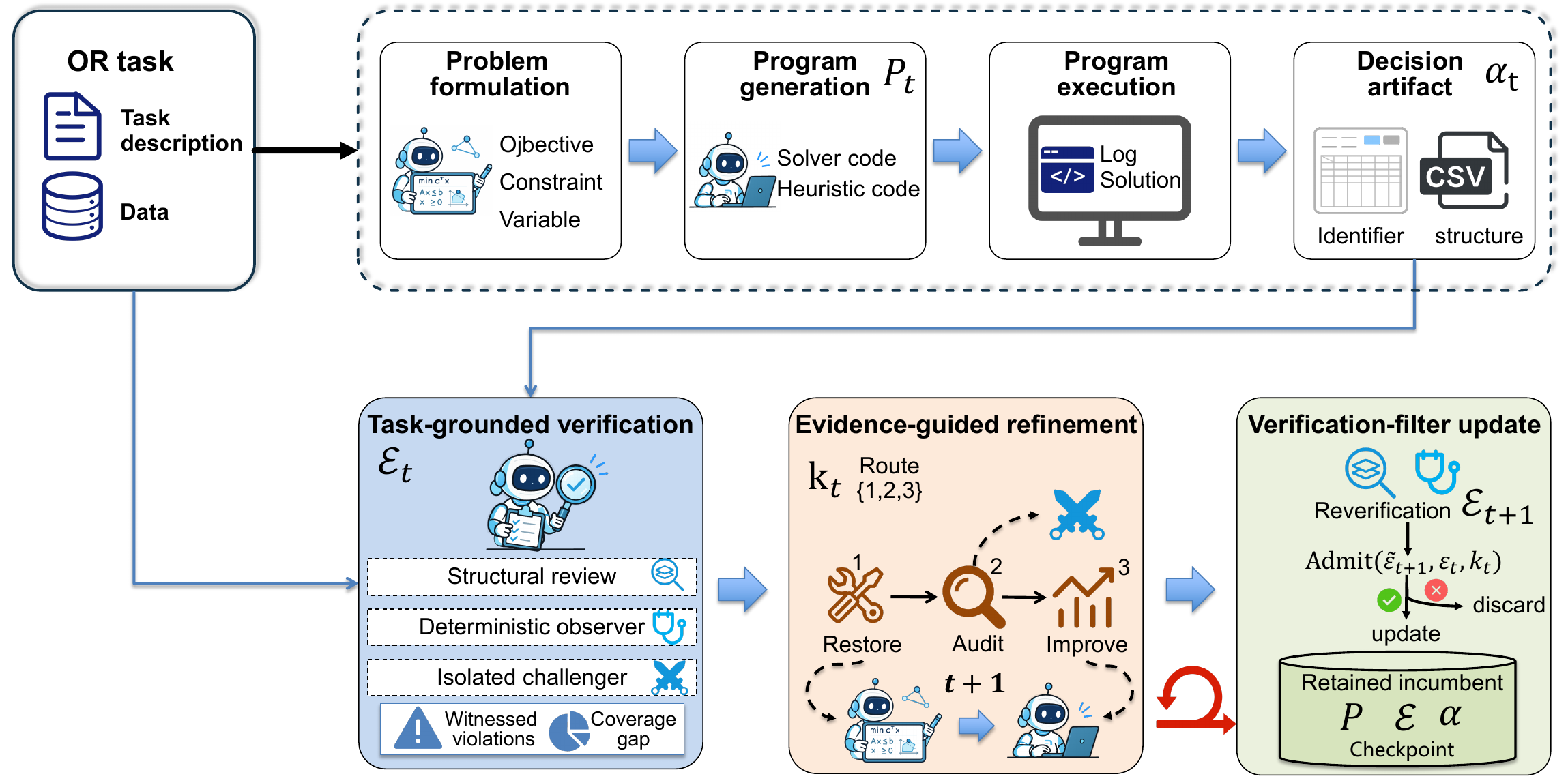}
\caption{Overview of the \method{} framework. A coding agent constructs and executes OR programs, while task-grounded verification guides conditional refinement, and verification-filtered updates retain accepted candidates.}
\label{fig:framework}
\end{figure*}

As illustrated in Figure~\ref{fig:framework}, a general-purpose coding agent formulates the problem, constructs an executable optimization program, and executes it to produce a decision artifact. VERAS organizes this process through three mechanisms. \emph{Task-Grounded Verification} establishes evidence of constraint violations and verification coverage. \emph{Evidence-Guided Refinement} determines subsequent solving actions based on this evidence. \emph{Verification-filtered Update} controls candidate acceptance to preserve established progress. At iteration $t$, the retained program $P_t$, its decision artifact $a_t$, and verification evidence $\mathcal E_t$ constitute the incumbent state.

\subsection{Task-Grounded Verification}
\label{sec:verification}

Task-grounded verification establishes evidence of candidate correctness by examining exported decisions against the original task requirements and instance data. For a decision artifact $a_t$, the verification process produces
\begin{equation}
\mathcal E_t =
\operatorname{Verify}(\mathcal I,a_t),
\label{eq:verification}
\end{equation}
which captures structural validity, witnessed constraint violations, and verification coverage.

\paragraph{Structural Review.}
The structural review establishes whether the decision artifact conforms to the task-specific output specification. It examines the required files, decision fields, data types, and completeness of the submitted records. Structural defects are reported with their affected fields or records, providing localized feedback for correcting missing or malformed outputs.

\paragraph{Deterministic Observer.}
The deterministic observer evaluates candidate decisions against task constraints reconstructed from the task requirements and instance data. For each supported constraint, the observer computes its satisfaction status and reports witnessed violations with affected decisions and numerical residuals where applicable. Each observer also records its verification scope, identifying the constraint families and conditions covered by its checks. These records distinguish observed violations from gaps in verification coverage and provide structured evidence for subsequent refinement.

\paragraph{Isolated Challenger.}
The isolated challenger extends verification by investigating requirements that remain insufficiently examined. Operating in a fresh context, it independently examines the task specification, instance data, and candidate decisions to search for diagnostic counterexamples. The resulting findings supplement deterministic observations and provide additional evidence for subsequent program correction.

\subsection{Evidence-Guided Refinement}
\label{sec:refinement}
Evidence-guided refinement employs rule-based diagnosis and ordered conditional routing to adapt subsequent solving actions to the candidate's verification state.

\paragraph{Evidence-Based Diagnosis.}
VERAS evaluates structural validity, witnessed constraint violations, and verification coverage to determine the next action. The controller conditionally activates three stages in a predefined order,
\begin{equation}
\begin{gathered}
\operatorname{Route}_k(\mathcal E,b)=
\begin{cases}
k, & g_k(\mathcal E,b)=1,\\
\mathsf{skip}, & \text{otherwise},
\end{cases}\\[3pt]
k:\ \mathsf{restore}\rightarrow
\mathsf{audit}\rightarrow
\mathsf{improve},
\end{gathered}
\label{eq:routing}
\end{equation}
where $g_k$ denotes the rule-based activation condition for stage $k$, determined by the latest verification evidence $\mathcal E$ and remaining budget $b$. Each stage is evaluated in order and skipped when its condition is unmet. Program revisions are reverified before subsequent decisions, while auditing augments the available verification evidence.

\paragraph{Feasibility Restoration.}
Structural defects or witnessed constraint violations trigger feasibility restoration. VERAS provides the coding agent with localized diagnostic evidence, including violated requirements and available residuals. The agent revises the mathematical formulation, heuristic strategy, or program implementation to address these defects. The revised program is executed and reverified before candidate admission.

\paragraph{Coverage-Aware Auditing.}
For viable candidates with incomplete verification coverage, VERAS invokes the isolated challenger described in Section~\ref{sec:verification}. Its findings supplement the verification evidence and provide diagnostic feedback for subsequent refinement. Auditing extends the available evidence without directly modifying the optimization program.

\paragraph{Performance Improvement.}
Viable candidates enter performance improvement when sufficient budget remains. The coding agent uses accumulated verification feedback to improve the original objective $f(\mathbf{x};\mathcal I)$ through formulation refinement, heuristic improvement, or additional optimization. Each proposed revision is executed and verified before candidate admission.

\subsection{Verification-filtered Update}
\label{sec:update}
Verification-filtered update controls candidate acceptance by reusing the structural review and deterministic observers from Section~\ref{sec:verification}.

\paragraph{Candidate Admission.}
For each proposed program $\widetilde P_{t+1}$, VERAS executes it to obtain a decision artifact $\widetilde a_{t+1}$ and collects verification evidence $\widetilde{\mathcal E}_{t+1}$. The admission decision is defined as
\begin{equation}
u_t = \operatorname{Admit}
(\widetilde{\mathcal E}_{t+1},\mathcal E_t,k_t)
\in \{0,1\},
\label{eq:admission}
\end{equation}
where $u_t=1$ indicates acceptance and $k_t$ denotes the current refinement phase. Admission requires non-regression in structural validity and supported constraint violations. Restoration permits incremental feasibility improvements, while performance-oriented updates additionally require valid artifacts without detected violations under supported checks.

\paragraph{Incumbent Retention.}
The incumbent is updated according to
\begin{equation}
\begin{aligned}
&(P_{t+1},a_{t+1},\mathcal E_{t+1})\\
&=
\begin{cases}
(\widetilde P_{t+1},\widetilde a_{t+1},
\widetilde{\mathcal E}_{t+1}), & u_t=1,\\
(P_t,a_t,\mathcal E_t), & u_t=0.
\end{cases}
\end{aligned}
\label{eq:incumbent}
\end{equation}
Accepted candidates replace the incumbent for subsequent refinement, while rejected candidates are discarded. The retained program and its decision artifact constitute the final output.
\section{Experimental Setup}
\label{sec:experiments}
\newcommand{\benchmodel}[2]{\raisebox{-0.14em}{\includegraphics[height=0.95em]{figs/model-icons/#1}}\hspace{0.35em}#2}

\begin{table*}[t]
\caption{Main results on ORAgentBench (107 tasks). Overall feasibility and quality are task-count-weighted across the easy, medium, and hard splits. Rows are grouped by harness and ordered by Overall pass rate within each group (ties are broken by Overall quality). \textbf{Bold} and \underline{underline} denote the best and second-best results in each column, respectively.}
\label{tab:oragent-main}
\centering
\footnotesize
\setlength{\tabcolsep}{3.2pt}
\renewcommand{\arraystretch}{1.10}
\resizebox{\textwidth}{!}{%
\begin{tabular}{@{}ll
>{\columncolor{veraseasybg}}c
>{\columncolor{verasmediumbg}}c
>{\columncolor{verashardbg}}c
c @{\hspace{5pt}}
>{\columncolor{veraseasybg}}c
>{\columncolor{verasmediumbg}}c
>{\columncolor{verashardbg}}c
c @{\hspace{5pt}}
>{\columncolor{veraseasybg}}c
>{\columncolor{verasmediumbg}}c
>{\columncolor{verashardbg}}c
c@{}}
\toprule
\multirow{2}{*}{\textbf{Harness}} & \multirow{2}{*}{\textbf{Model}} &
\multicolumn{4}{c}{\textbf{Feasibility $\uparrow$}} &
\multicolumn{4}{c}{\textbf{Quality $\uparrow$}} &
\multicolumn{4}{c}{\textbf{Pass rate (\%) $\uparrow$}} \\
\cmidrule(lr){3-6}\cmidrule(lr){7-10}\cmidrule(l){11-14}
& & Easy & Medium & Hard & Overall & Easy & Medium & Hard & Overall & Easy & Medium & Hard & Overall \\
\midrule
\multirow{3}{*}{Codex}
& \benchmodel{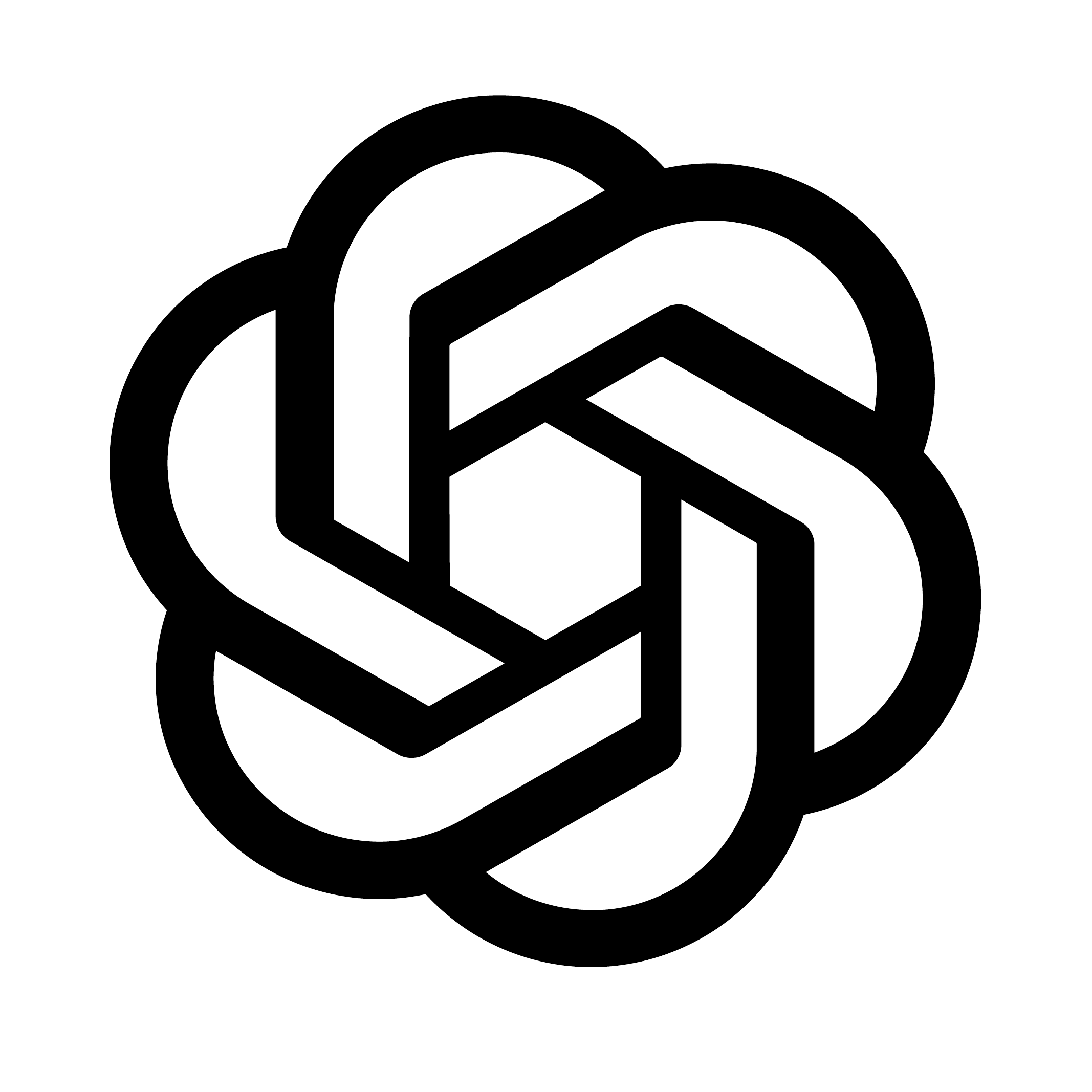}{GPT-5.4}       & \underline{0.69} & \textbf{0.68} & \underline{0.32} & \underline{0.57} & 0.45 & 0.31 & \underline{0.21} & \underline{0.32} & 46.88 & \underline{39.02} & \underline{20.59} & \underline{35.51} \\
& \benchmodel{openai.pdf}{GPT-5.3-Codex} & 0.50 & \textbf{0.68} & \underline{0.32} & 0.51 & 0.30 & \textbf{0.37} & 0.19 & 0.29 & 31.25 & \textbf{43.90} & \underline{20.59} & 32.71 \\
& \benchmodel{openai.pdf}{GPT-5.4-mini}  & 0.62 & 0.54 & 0.21 & 0.46 & 0.43 & 0.24 & 0.07 & 0.24 & 43.75 & 29.27 & 5.88 & 26.17 \\
\midrule
\multirow{6}{*}{Claude Code}
& \benchmodel{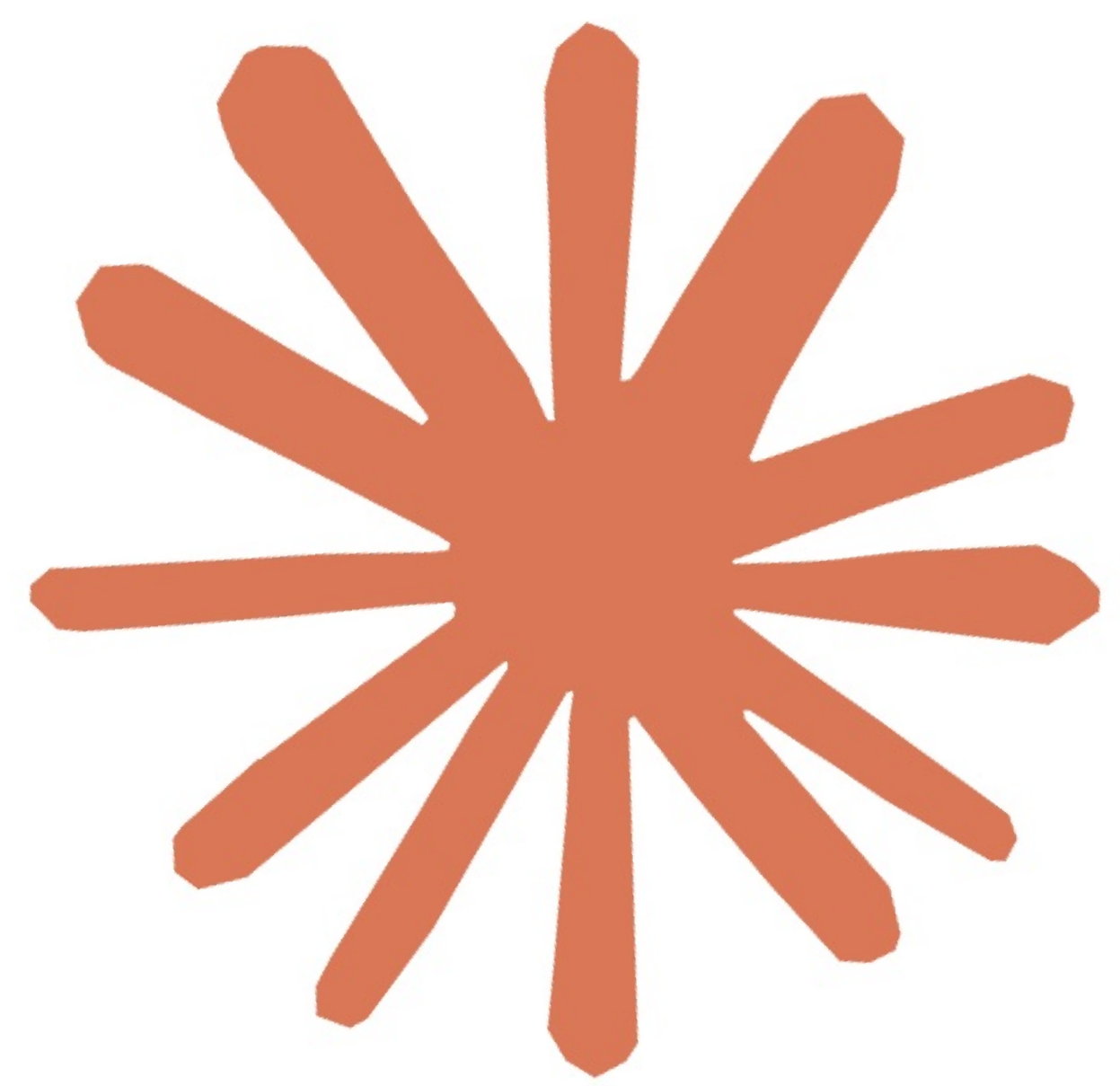}{Claude Opus 4.6}   & 0.66 & \underline{0.59} & 0.21 & 0.49 & \underline{0.49} & 0.32 & 0.14 & 0.31 & \underline{50.00} & \underline{39.02} & 14.71 & 34.58 \\
& \benchmodel{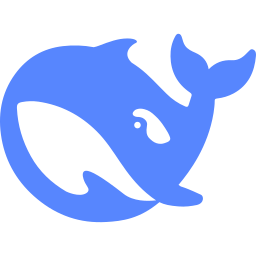}{DeepSeek V4 Pro}   & 0.55 & 0.46 & 0.23 & 0.41 & 0.38 & 0.25 & 0.14 & 0.25 & 39.58 & 28.46 & 13.73 & 27.10 \\
& \benchmodel{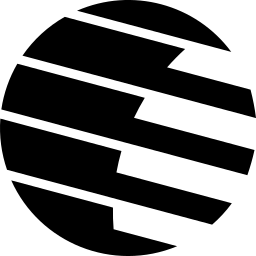}{Kimi K2.6}              & 0.50 & 0.51 & 0.26 & 0.43 & 0.29 & 0.25 & 0.13 & 0.22 & 31.25 & 36.59 & 11.76 & 27.10 \\
& \benchmodel{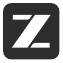}{GLM 5.1}                 & 0.53 & 0.46 & 0.15 & 0.38 & 0.37 & 0.29 & 0.10 & 0.25 & 37.50 & 31.71 & 8.82 & 26.17 \\
& \benchmodel{deepseek.png}{DeepSeek V4 Flash} & 0.62 & 0.41 & 0.15 & 0.39 & 0.45 & 0.21 & 0.09 & 0.24 & 46.88 & 26.83 & 5.88 & 26.17 \\
& \benchmodel{claude.pdf}{Claude Sonnet 4.6}    & 0.47 & 0.51 & 0.26 & 0.42 & 0.31 & 0.23 & 0.14 & 0.23 & 31.25 & 29.27 & 14.71 & 25.23 \\
\midrule
\textbf{\method{}} & \textbf{\benchmodel{openai.pdf}{GPT-5.4}} & \textbf{0.84} & \textbf{0.68} & \textbf{0.50} & \textbf{0.67} & \textbf{0.55} & \underline{0.34} & \textbf{0.31} & \textbf{0.39} & \textbf{56.25} & \underline{39.02} & \textbf{35.29} & \textbf{42.99} \\
\bottomrule
\end{tabular}%
}
\end{table*}

We evaluate VERAS on ORAgentBench and two industrial-scale OR problems to assess its reliability and optimization performance across heterogeneous tasks. ORAgentBench evaluates end-to-end problem solving, while the industrial studies examine heuristic inter-datacenter placement and mathematical programming for active distribution network planning.

\subsection{Benchmark Study}
\label{sec:benchmark-setup}
\paragraph{Task.}
ORAgentBench~\citep{li2026oragentbench} comprises 107 OR tasks (32 easy, 41 medium, and 34 hard instances) across eight application families, spanning static, stochastic, robust, and multi-stage optimization. Agents construct executable optimization programs and submit complete decision artifacts from task requirements and instance data. Multi-stage tasks additionally require later decisions to respect previously committed state; the benchmark reveals the next event only after the preceding stage meets its reward gate. Evaluation concerns the exported decisions, rather than the plausibility of the generated formulation alone.

\paragraph{Agent configuration.}
We use GPT-5.4 with high reasoning effort and two concurrent task workers. Each worker operates in an isolated environment with Python, PySCIPOpt, and SCIP. The task-stage wall-clock ceiling is 2,700 seconds; candidate SCIP calls have a 300-second ceiling and a target relative gap of $5\times10^{-4}$. The controller reserves a fraction of 0.30 for recovery, with a minimum recovery allocation of 120 seconds, and allocates a fraction of 0.22 of the remaining budget to the challenger. These are conditional allocations, not fixed numbers of refinement rounds. Appendix~\ref{app:hyperparameters} lists the remaining runtime settings.

\paragraph{Verification and scoring.}
Agents receive the public statement, instance data, and output templates, but not the benchmark evaluator or reference artifacts. Structural checks and deterministic observers inspect exported decisions against supported public constraints; the challenger supplies additional diagnostic evidence. These online checks guide refinement but do not replace the external evaluator. For task $i$, let $F_i$ and $Q_i$ denote its externally evaluated feasibility and normalized quality. We report
\begin{equation}
\mathrm{PassRate}=\frac{100}{107}\sum_{i=1}^{107}
\mathbb{I}[F_i>0\;\land\;Q_i>0.4].
\label{eq:passrate}
\end{equation}
Quality follows the benchmark's task-specific normalization against a verified reference objective and, when available, a valid global bound. It is distinct from both the raw objective and the solver's reported optimality gap. Successful program execution alone does not establish constraint feasibility or adequate normalized objective quality. Incomplete or infeasible submissions receive zero credit. Overall feasibility and quality are means over tasks, equivalent to weighting the three difficulty splits by 32, 41, and 34; they are not unweighted means of the split averages. For multi-stage tasks, the aggregation unit is the complete required trajectory, not an individual successful stage.

\paragraph{Comparisons.}
The Codex and Claude Code rows in Table~\ref{tab:oragent-main} are published benchmark results, not new baseline runs. They use the same official scoring criteria, but this comparison does not control realized inference compute. The six-Hard-task ablation uses the same per-stage limits and evaluation protocol, retains the original VERAS results, and evaluates component variants in separate runs (Section~\ref{sec:ablations}). It is a diagnostic comparison rather than a repeated-run estimate of each component's effect. Benchmark outcomes informed harness development; the full suite therefore serves as a development benchmark.

\subsection{Case Study: Inter-Datacenter Bandwidth Optimization}
\paragraph{Task.}
This study optimizes the placement of 625,191 tasks across Zhongwei and Huailai, China. Task assignments determine table locations and cross-region transfers. The objective is to minimize peak inter-datacenter bandwidth subject to regional vCore and memory capacities; migration volume is an auxiliary operational metric. Each candidate is an executable placement program, not only a saved task-to-region mapping.

\paragraph{Protocol.}
We compare Operational Mapping, the Historical-Majority Heuristic, native Codex, and VERAS. Operational Mapping preserves the archived assignments. The historical heuristic assigns each task to its most frequently observed execution region, defaulting to Zhongwei when no history is available. All methods use the same placement constraints and bandwidth metrics. Agent-generated programs must regenerate their assignments within the 120-second dispatcher limit; this execution limit is distinct from the agent's search budget.

Industrial task inputs are agent-visible but are not redistributed; the source-replay results refer to checks performed with those inputs available (Appendix~\ref{app:reproducibility}). The case uses August 2025 traces. August 8--9 records are available for program construction, while August 10 event records are excluded from agent inputs and used for evaluation. Regional resource limits are fixed constants derived from capacity peaks over August 7--10. Both agent systems use GPT-5.6 Sol with xhigh reasoning effort, a timeout multiplier of three, and a 19,800-second runtime limit. The comparison pairs an archived VERAS run with a later native Codex run; matching hashes establish identical task inputs and evaluator versions. Earlier development inspected August 10 scores, so this is a retrospective comparison, not an untouched temporal test.

\subsection{Case Study: Active Distribution Network Planning}
\paragraph{Task.}
This study addresses mixed-integer second-order-cone planning for a 54-bus active distribution network with 63 branches, four scenarios, and 24 hourly periods. The objective minimizes annualized investment, operation, and maintenance costs through joint infrastructure and operating decisions under electrical and operational constraints. Agents construct executable optimization programs that export complete planning and operational decisions.

\paragraph{Protocol.}
We compare two expert-designed models, Hourly MISOCP and Varying-Timestep MISOCP, with native Codex and VERAS on the same physical instance. Both agent systems use GPT-5.6 Sol with xhigh reasoning effort, identical task inputs, and the same independent evaluation protocol. Solver execution is limited to four threads and 900 seconds. The VERAS profile has a 15,360-second agent wall-time limit and a timeout multiplier of two. The reported VERAS solution is a separately selected run, rather than the VERAS arm executed alongside native Codex.

Both reported agent solutions are regenerated from the agent-visible task data using a stricter 750-second solver limit. Replay time includes setup and artifact generation, so it can exceed 750 seconds without violating the solver limit. Replaying a fixed program tests artifact reproducibility; it is not a repetition of the full agent search. Independent audits recompute constraints and objectives from the exported decisions.

The Hourly MISOCP model uses MATLAB/YALMIP with MOSEK. It jointly optimizes line and substation investments, discrete storage/PV/wind capacities, and hourly operations across four scenarios. In the fault scenario, branches 34 and 44 are disconnected during hours 7--8, with reduced substation capacities. Varying-Timestep MISOCP adds five-minute storage charging, discharging, and state-of-charge decisions during this window, coupled to hourly decisions through boundary states, energy consistency, and power headroom. It is therefore a finer-grained operating formulation, not merely another solver configuration for the hourly model.

\section{Results and Discussion}

\subsection{Benchmark Study Results} 
On ORAgentBench, VERAS passes 46 of 107 tasks (42.99\%), exceeding the published GPT-5.4 Codex baseline of 38 tasks (35.51\%) by 7.48 percentage points under the same official evaluation criteria (Table~\ref{tab:oragent-main}). The largest improvement occurs on Hard tasks, where the pass count increases from 7/34 to 12/34, while Medium performance remains unchanged at 16/41. Mean feasibility increases from 0.57 to 0.67, accompanied by an improvement in normalized solution quality from 0.32 to 0.39.

Execution trajectories further illustrate how task-grounded verification supports program refinement. Across 117 task stages, VERAS records 15 structural restorations and three reductions in witnessed constraint residuals. In online electric-bus depot replanning, deterministic observers identify 42 trip-row nullability violations, all resolved through restoration. The controller also invokes isolated challengers for 103 tasks and admits 114 of 132 performance revisions, demonstrating the integration of independent auditing and verification-filtered updates into the solving process.

Hard tasks remain challenging, with 17/34 producing feasible solutions but only 12/34 satisfying the solution-quality threshold. In addition, four tasks fail to complete all required stages. These results highlight the distinct challenges of constraint feasibility, solution quality, and multi-stage completion, motivating the controlled ablation study that follows.

\subsection{Ablation Study}
We conduct an ablation study on six Hard tasks using GPT-5.4, comparing VERAS against four variants that disable the isolated challenger, deterministic observers, feasibility restoration routing, and verification-filtered admission, respectively.
VERAS uses frozen results from the original evaluation, while the four ablations are evaluated in separate runs, with API-affected trials replaced based on run validity.

\begin{figure}[t]
    \centering
    \includegraphics[width=\linewidth]{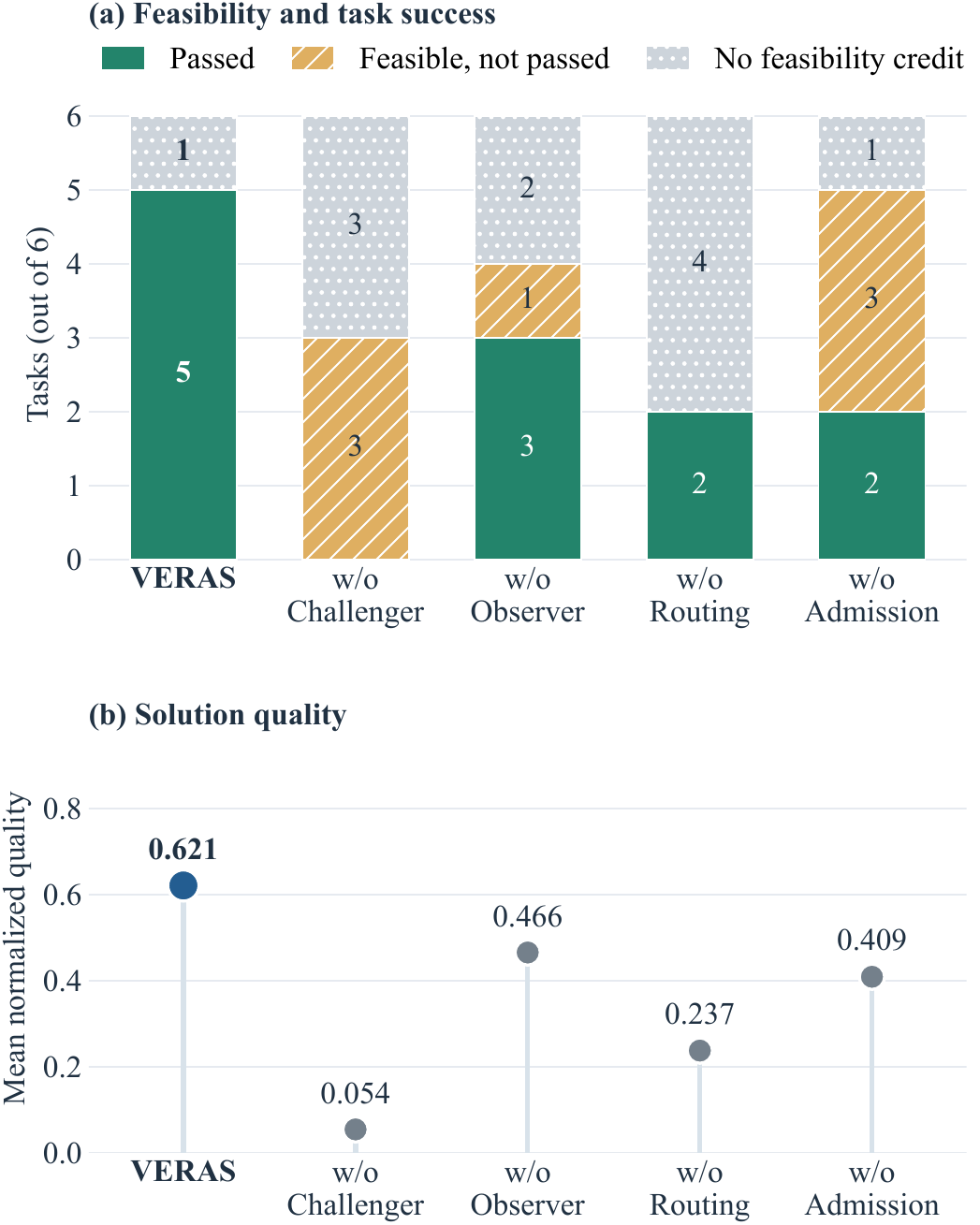}
    \caption{
     Ablation study on six Hard tasks.
    (a) Feasibility and pass outcomes across variates.
    (b) Mean normalized solution quality.
    VERAS reports the frozen original evaluation results.
    }
    \label{fig:ablation}
\end{figure}

Figure~\ref{fig:ablation} shows that VERAS achieves the highest task success and solution quality, passing 5/6 tasks with a mean normalized quality of 0.621. Removing the isolated challenger reduces the pass count from five to zero and quality to 0.054, highlighting the importance of additional constraint auditing. Disabling the deterministic observer reduces feasible tasks from five to four, passes from five to three, and quality to 0.466. Without feasibility restoration routing, only two tasks remain feasible and pass, with quality dropping to 0.237. Finally, removing verification-filtered admission preserves five feasible tasks but reduces the pass count to two and quality to 0.409, indicating its role in retaining higher-quality solutions. Together, these results demonstrate the complementary contributions of task-grounded verification, evidence-guided refinement, and verification-filtered updates.

% \begin{table}[!ht]
% \color{verasred}
% \caption{\textcolor{verasred}{Matched-budget six-task Hard diagnostic ablation. The VERAS row reports the frozen result from the original Hard evaluation; the remaining rows report the completed component-ablation evaluations described in the text. Tokens include input (including cached input) and output tokens.}}
% \label{tab:ablations}
% \centering
% \small
% \setlength{\tabcolsep}{6pt}
% \resizebox{\linewidth}{!}{%
% \begin{tabular}{@{}lccccc@{}}
% \toprule
% Configuration & Feasible (/6) $\uparrow$ & Passed (/6) $\uparrow$ & Mean quality $\uparrow$ & Tokens/task $\downarrow$ & Wall time/task $\downarrow$\\
% \midrule
% VERAS & 5 & 5 & 0.6214 & 6.70M & 27.8 min\\
% No challenger & 3 & 0 & 0.0539 & 4.73M & 25.2 min\\
% No semantic observer & 4 & 3 & 0.4655 & 8.20M & 31.4 min\\
% No restoration routing & 2 & 2 & 0.2373 & 1.96M & 18.5 min\\
% No admission filter & 5 & 2 & 0.4090 & 8.28M & 30.5 min\\
% \bottomrule
% \end{tabular}
% }
% \end{table}

% \textcolor{verasred}{The saved 107-task trajectories contain no naturally triggered checkpoint rollback, so we do not present a natural-performance ``no rollback'' row. Rollback can instead be evaluated with a separate controlled-corruption test that measures recovery behavior without additional model calls. For this ablation, we report final feasibility, pass, normalized quality, tokens, wall time, and mechanism-event counts from the saved event streams.}

\subsection{Case Study: Inter-datacenter placement}
Table~\ref{tab:bandwidth} shows that VERAS achieves the lowest peak inter-datacenter bandwidth of 1,110.603 Gbps, reducing the peak by 10.34\% compared with native Codex (1,238.625 Gbps). Both agent-generated programs independently reproduce their respective archived placement outputs byte-for-byte from public inputs within the 120-second dispatcher limit. The expert-designed mappings exhibit substantially higher bandwidth peaks, ranging from 18,276.071 to 19,495.167 Gbps. VERAS incurs slightly higher migration than native Codex (25.901 versus 25.473 PB), reflecting a trade-off between bandwidth reduction and data movement, with migration treated as an auxiliary metric.

The comparison uses an archived VERAS result and a subsequent native Codex run. A separate VERAS run achieves 1,567.303 Gbps (Appendix~\ref{app:dc-replicate}), illustrating variability across optimization runs.

\begin{table}[!ht]
\caption{Inter-datacenter placement results. Lower peak bandwidth and migration are better.}
\label{tab:bandwidth}
\centering
\small
\resizebox{\linewidth}{!}{%
\begin{tabular}{@{}lrr@{}}
\toprule
Method & Peak (Gbps) $\downarrow$ & Migration (PB) $\downarrow$\\
\midrule
Operational mapping          & 19,495.167 & 0.000\\
Historical-majority heuristic & 18,276.071 & 0.333\\
Native Codex                 & 1,238.625  & 25.473\\
\textbf{VERAS}               & \textbf{1,110.603} & 25.901\\
\bottomrule
\end{tabular}%
}
\end{table}
% On the August 10 replay, VERAS attains a strictly feasible peak of 1,110.603 Gbps, compared with 1,238.625 Gbps for native Codex and 19,495.167 Gbps for the expert operational mapping (Table~\ref{tab:bandwidth}). This is 10.336\% below native Codex. Both agent-generated mappings replay byte-identically from public inputs within the 120-second dispatcher limit. Migration is an auxiliary metric.

% \begin{table}[!ht]
% \caption{\revision{Inter-datacenter placement on August 10. Expert mappings are human baselines; lower peak bandwidth and migration are better.}}
% \label{tab:bandwidth}
% \centering
% \small
% % \setlength{\tabcolsep}{5.5pt}
% \resizebox{\linewidth}{!}{%
% \begin{tabular}{@{}lrr@{}}
% \toprule
% \revision{Method} & \revision{Peak (Gbps) $\downarrow$} & \revision{Migration (PB) $\downarrow$}\\
% \midrule
% \revision{Operational mapping} & \revision{19,495.167} & \revision{0.000}\\
% \revision{Historical-majority heuristic} & \revision{18,276.071} & \revision{0.333}\\
% \revision{Native Codex} & \revision{1,238.625} & \revision{25.473}\\
% \revision{\textbf{VERAS}} & \revision{\textbf{1,110.603}} & \revision{25.901}\\
% \bottomrule
% \end{tabular}%
% }
% \end{table}

% The VERAS candidate was archived before the later native Codex run; this comparison is retrospective. A fresh VERAS repeat yields 1,567.303 Gbps (Appendix~\ref{app:dc-replicate}).

\subsection{Case Study: 54-bus Active distribution network}
Table~\ref{tab:adn} shows that VERAS achieves the lowest annualized total cost of USD~39.424M/year, reducing costs by 6.28\% and 13.71\% compared with the expert-designed Hourly MISOCP and native Codex, respectively. The improvement over the expert model is primarily driven by lower investment costs (USD~16.154M versus USD~18.984M), despite a slightly higher operating cost (USD~23.269M versus USD~23.081M). Compared with native Codex, VERAS reduces both investment and operating costs. Both agent-generated solutions pass independent constraint and objective audits and can be regenerated from public inputs.

\begin{table}[!ht]
\caption{Annualized costs for 54-bus active distribution network planning and fault restoration (million USD/year). Lower is better.}
\label{tab:adn}
\centering
\small
\setlength{\tabcolsep}{4.2pt}
\resizebox{\linewidth}{!}{%
\begin{tabular}{@{}lrrr@{}}
\toprule
Method & Total $\downarrow$ & Investment $\downarrow$ & Operation $\downarrow$\\
\midrule
Hourly MISOCP
    & 42.065 & 18.984 & \textbf{23.081}\\
Varying-timestep MISOCP
    & 42.923 & 19.012 & 23.911\\
Native Codex
    & 45.690 & 18.490 & 27.200\\
\textbf{VERAS}
    & \textbf{39.424} & \textbf{16.154} & 23.269\\
\bottomrule
\end{tabular}%
}
\end{table}
\FloatBarrier

% \revision{VERAS attains an independently audited annualized cost of USD~39.424M, below the expert hourly model (USD~42.065M), its varying-timestep extension (USD~42.923M), and native Codex (USD~45.690M) on the same 54-bus instance (Table~\ref{tab:adn}). The reductions are 6.279\% relative to the expert hourly model and 13.715\% relative to native Codex. The latter is a comparison across separate agent runs, not a contemporaneous pair. Both agent solutions pass independent constraint and objective audits and regenerate from public source.}

% \begin{table}[!ht]
% \caption{\revision{54-bus planning and fault restoration. Annualized cost is in million USD/year; lower is better.}}
% \label{tab:adn}
% \centering
% \small
% \setlength{\tabcolsep}{4.2pt}
% \resizebox{\linewidth}{!}{%
% \begin{tabular}{@{}lrrr@{}}
% \toprule
% \revision{Method} & \revision{Total $\downarrow$} & \revision{Investment} & \revision{Operation}\\
% \midrule
% \revision{Hourly MISOCP} & \revision{42.065} & \revision{18.984} & \revision{23.081}\\
% \revision{Varying-timestep MISOCP} & \revision{42.923} & \revision{19.012} & \revision{23.911}\\
% \revision{Native Codex} & \revision{45.690} & \revision{18.490} & \revision{27.200}\\
% \revision{\textbf{VERAS}} & \revision{\textbf{39.424}} & \revision{\textbf{16.154}} & \revision{23.269}\\
% \bottomrule
% \end{tabular}%
% }
% \end{table}
% \FloatBarrier

\section{Conclusion}
We presented VERAS, a verification-guided adaptive search harness for reliable operations research algorithm discovery. VERAS integrates task-grounded verification, evidence-guided refinement, and verification-filtered updates to identify constraint violations, distinguish gaps in verification coverage, and guide subsequent program improvement while preserving established verification evidence. Experiments on ORAgentBench and two industrial optimization problems demonstrate its effectiveness across mathematical programming and heuristic algorithm design. These findings highlight the importance of moving beyond pass-or-fail evaluation toward evidence-driven search, providing a practical direction for developing more reliable and effective OR agents.

\section*{Limitations}
\paragraph{Evaluation scope.}
ORAgentBench informed harness development, including inspection of task-level failures. The reported aggregate is therefore performance on a development benchmark, not an unbiased estimate for unseen tasks or optimization domains. Published baselines use the same scoring criteria but were not rerun as part of this study. The six-task ablation uses separate runs and a frozen VERAS reference, and is too small to establish reliable component-level effect sizes. Shared time ceilings also do not imply equal realized inference compute.

\paragraph{Application comparisons.}
The industrial studies cover one datacenter setting and one fixed distribution-network instance. The datacenter comparison is retrospective: August 10 records are withheld within runs, but earlier development inspected scores from that date, and resource limits use a capacity envelope that includes it. The additional VERAS run does not preserve the archived run's advantage over native Codex. The grid comparison likewise reports a separately selected VERAS run, while the expert models differ in temporal resolution. These studies demonstrate the reported programs' outcomes, rather than an average advantage under repeated, prospectively matched experiments.

\paragraph{Verification coverage.}
Deterministic observers check only implemented constraint families. A zero residual establishes consistency with those checks, not correctness with respect to every task requirement. A fresh challenger context reduces shared execution history but does not eliminate model-level biases or guarantee that omitted constraints will be found. Admission filtering preserves properties covered by the checks; it does not establish monotonic objective improvement or global optimality. Generated optimization programs still require domain-specific validation before operational use.

\paragraph{Reproducibility.}
The experiments rely on proprietary model services, and the reported source replays reproduce fixed programs rather than entire agent trajectories. The released industrial summaries and code do not include the original operational input rows or generated decision archives, which limits independent reproduction of the application results. The current comparisons also do not establish cost-effectiveness under a common token or monetary budget.

\bibliography{references}

@article{romeraparedes2024funsearch,
  title={Mathematical discoveries from program search with large language models},
  author={Romera-Paredes, Bernardino and Barekatain, Mohammadamin and Novikov, Alexander and Balog, Matej and Kumar, M. Pawan and Dupont, Emilien and Ruiz, Francisco J. R. and Ellenberg, Jordan S. and Wang, Pengming and Fawzi, Omar and Kohli, Pushmeet and Fawzi, Alhussein},
  journal={Nature},
  volume={625},
  pages={468--475},
  year={2024},
  doi={10.1038/s41586-023-06924-6}
}

@article{song2026nemo,
  title={NEMO: Execution-Aware Optimization Modeling via Autonomous Coding Agents},
  author={Song, Yang and Vyas, Anoushka and Wei, Zirui and Pakazad, Sina Khoshfetrat and Ohlsson, Henrik and Neubig, Graham},
  journal={arXiv preprint arXiv:2601.21372},
  year={2026}
}

@inproceedings{xiao2024chain,
  title={{Chain-of-Experts}: When {LLMs} Meet Complex Operations Research Problems},
  author={Xiao, Ziyang and Zhang, Dongxiang and Wu, Yangjun and Xu, Lilin and Wang, Yuan and Han, Xiongwei and Fu, Xiaojin and Zhong, Tao and Zeng, Jia and Song, Mingli and Chen, Gang},
  booktitle={International Conference on Learning Representations},
  volume={2024},
  pages={48519--48537},
  year={2024},
}

@article{an2025repositioning,
  title={Repositioning coal power to accelerate net-zero transition of China’s power system},
  author={An, Kangxin and Zheng, Xinzhu and Shen, Jianxiang and Xie, Canyang and Wang, Can and Cai, Wenjia and Bu, Chujie},
  journal={Nature Communications},
  volume={16},
  number={1},
  pages={2311},
  year={2025},
  doi={10.1038/s41467-025-57559-2}
}

@article{yao2022react,
  title={React: Synergizing reasoning and acting in language models},
  author={Yao, Shunyu and Zhao, Jeffrey and Yu, Dian and Du, Nan and Shafran, Izhak and Narasimhan, Karthik and Cao, Yuan},
  journal={arXiv preprint arXiv:2210.03629},
  year={2022},
}

@article{guo2025deepseek,
  title={DeepSeek-R1 incentivizes reasoning in LLMs through reinforcement learning},
  author={Guo, Daya and Yang, Dejian and Zhang, Haowei and Song, Junxiao and Wang, Peiyi and Zhu, Qihao and Xu, Runxin and Zhang, Ruoyu and Ma, Shirong and Bi, Xiao and others},
  journal={Nature},
  volume={645},
  number={8081},
  pages={633--638},
  year={2025},
  doi={10.1038/s41586-025-09422-z}
}

@book{williams2013model,
  title={Model Building in Mathematical Programming},
  author={Williams, H. Paul},
  edition={5},
  year={2013},
  publisher={John Wiley \& Sons},
  isbn={9781118443330}
}

@article{colangelo2026ai,
  title={{AI} data centres as grid-interactive assets},
  author={Colangelo, Philip and Coskun, Ayse K. and Megrue, Jack and Roberts, Ciaran and Sengupta, Shayan and Sivaram, Varun and Tiao, Ethan and Vijaykar, Aroon and Williams, Chris and Wilson, Daniel C. and Records, Brandon and MacFarland, Zack and Dreiling, Daniel and Morey, Nathan and Ratnayake, Anuja and Vairamohan, Baskar},
  journal={Nature Energy},
  volume={11},
  number={2},
  pages={254--261},
  year={2026},
  doi={10.1038/s41560-025-01927-1}
}

@article{vazifeh2018addressing,
  title={Addressing the minimum fleet problem in on-demand urban mobility},
  author={Vazifeh, Mohammad M and Santi, Paolo and Resta, Giovanni and Strogatz, Steven H and Ratti, Carlo},
  journal={Nature},
  volume={557},
  number={7706},
  pages={534--538},
  year={2018},
  doi={10.1038/s41586-018-0095-1}
}

@article{novikov2025alphaevolve,
  title={{AlphaEvolve}: A coding agent for scientific and algorithmic discovery},
  author={Novikov, Alexander and V\~u, Ng\^an and Eisenberger, Marvin and Dupont, Emilien and Huang, Po-Sen and Wagner, Adam Zsolt and Shirobokov, Sergey and Kozlovskii, Borislav and Ruiz, Francisco J. R. and Mehrabian, Abbas and Kumar, M. Pawan and See, Abigail and Chaudhuri, Swarat and Holland, George and Davies, Alex and Nowozin, Sebastian and Kohli, Pushmeet and Balog, Matej},
  journal={arXiv preprint arXiv:2506.13131},
  year={2025}
}

@inproceedings{ahmaditeshnizi2024optimus,
  title={{OptiMUS}: Scalable Optimization Modeling with ({MI}){LP} Solvers and Large Language Models},
  author={Ahmaditeshnizi, Ali and Gao, Wenzhi and Udell, Madeleine},
  booktitle={Proceedings of the 41st International Conference on Machine Learning},
  series={Proceedings of Machine Learning Research},
  volume={235},
  pages={577--596},
  year={2024},
  publisher={PMLR},
}

@article{huang2024orlm,
  title={{ORLM}: A customizable framework in training large models for automated optimization modeling},
  author={Huang, Chenyu and Tang, Zhengyang and Hu, Shixi and Jiang, Ruoqing and Zheng, Xin and Ge, Dongdong and Wang, Benyou and Wang, Zizhuo},
  journal={arXiv preprint arXiv:2405.17743},
  year={2024}
}

@article{zhang2025orllmagent,
  title={{OR-LLM-Agent}: Automating Modeling and Solving of Operations Research Optimization Problems with Reasoning {LLM}},
  author={Zhang, Bowen and Luo, Pengcheng and Yang, Genke and Soong, Boon-Hee and Yuen, Chau},
  journal={arXiv preprint arXiv:2503.10009},
  year={2025}
}

@article{li2026oragentbench,
  title={{ORAgentBench}: Can {LLM} agents solve challenging operations research tasks end to end?},
  author={Li, Jiajun and Cai, Mingshu and Li, Yixuan and Ding, Yu and Hou, Ran and Nie, Guanyu and Han, Xiongwei and Wang, Wanyuan},
  journal={arXiv preprint arXiv:2606.19787},
  year={2026}
}

@article{fang2026trival,
  title={{TriVAL}: A tri-validation framework for faithful automatic optimization modeling},
  author={Fang, Ziyang and Wang, JinXi and Zhong, Jinghui and Ong, Yew-Soon},
  journal={arXiv preprint arXiv:2605.23966},
  year={2026},
}

@inproceedings{liu2024eoh,
  title={Evolution of Heuristics: Towards Efficient Automatic Algorithm Design Using Large Language Model},
  author={Liu, Fei and Tong, Xialiang and Yuan, Mingxuan and Lin, Xi and Luo, Fu and Wang, Zhenkun and Lu, Zhichao and Zhang, Qingfu},
  booktitle={Proceedings of the 41st International Conference on Machine Learning},
  series={Proceedings of Machine Learning Research},
  volume={235},
  pages={32201--32223},
  year={2024},
  publisher={PMLR},
}

@inproceedings{ye2024reevo,
  title={{ReEvo}: Large Language Models as Hyper-Heuristics with Reflective Evolution},
  author={Ye, Haoran and Wang, Jiarui and Cao, Zhiguang and Berto, Federico and Hua, Chuanbo and Kim, Haeyeon and Park, Jinkyoo and Song, Guojie},
  booktitle={Advances in Neural Information Processing Systems},
  volume={37},
  pages={43571--43608},
  year={2024},
  doi={10.52202/079017-1381}
}

@article{liu2026optiver,
  title={{Opt-Verifier}: Unleashing the Power of {LLMs} for Optimization Modeling via Dual-Side Verification},
  author={Liu, Haoyang and Wang, Jie and Niu, Boxuan and Han, Xiongwei and Xu, Yian and Ye, Mingxuan and Geng, Zijie and Zhu, Fangzhou and Zhong, Tao and Yuan, Mingxuan and Hao, Jianye},
  journal={arXiv preprint arXiv:2605.29556},
  year={2026},
}
\clearpage
\appendix
\section{Extended Experimental Details}
\label{app:experimental-details}

\subsection{Information Available During Evaluation}
\label{app:information-boundary}

The controller revises candidates using the task materials, exported decisions, and online verification reports. These reports are distinct from benchmark scores. Table~\ref{tab:app-information-boundary} lists the inputs and outputs of each role. In multi-stage benchmark tasks, a stage-level reward gate determines whether the next event is revealed~\citep{li2026oragentbench}; this does not expose the hidden evaluator code or reference solutions to the agent.

\begin{table}[!ht]
\caption{Inputs and outputs of the solving and evaluation roles. Evaluation scores a submitted artifact; it is not a replacement for the controller's online checks.}
\label{tab:app-information-boundary}
\centering
\small
\setlength{\tabcolsep}{4pt}
\renewcommand{\arraystretch}{1.10}
\begin{tabularx}{\linewidth}{@{}p{0.24\linewidth}X@{}}
\toprule
Role & Input and output \\
\midrule
Producer & Task packet, prior-stage artifacts, and refinement feedback; executable program and decisions \\
Observer & Task inputs and exported artifact; structural checks, constraint residuals, and declared coverage \\
Challenger & Task packet, candidate artifact, and observer evidence; read-only search for counterexamples \\
Evaluator & Submitted artifact and evaluation instance; feasibility and quality scores \\
\bottomrule
\end{tabularx}
\end{table}

On ORAgentBench, agents receive the public statement, instance data, and output templates, together with Python, PySCIPOpt, and SCIP. Deterministic observers check supported constraints on exported decisions and record their scope. A zero residual refers only to that scope. For a viable candidate with incomplete coverage and sufficient remaining time, the controller can request a read-only challenger review. Final task outcomes require completion of all mandatory stages.

In the datacenter study, August 8--9, 2025 transfer records are visible to the agents, but August 10 event records are not. Resource limits are fixed constants derived from the August 7--10 capacity envelope. Table~\ref{tab:bandwidth} compares an archived VERAS run with a later native Codex run using matched inputs and evaluation code. The archive was selected after earlier August 10 feedback. Thus, withholding event records within a run does not make the development process independent of that evaluation date.

For both industrial studies, \emph{agent-visible inputs} refers to data supplied to the agents during the experiment, not necessarily data available in the public release. Release restrictions are described in Appendix~\ref{app:reproducibility}.

\subsection{Prompt Construction}
\label{app:prompts}

The initial prompt combines the task instruction, a runtime note, and deterministic guidance selected by public task family. Subsequent prompts are conditional on the controller state and include serialized structural audits and constraint observations. The released harness contains the prompt text; the descriptions below summarize its role in the search procedure.

\paragraph{Feasibility restoration.}
When an artifact is missing, structurally invalid, or violates an observed constraint, the restoration prompt supplies the corresponding audit. It prioritizes feasibility and specifies an elastic Phase-I formulation when the exact model has no incumbent. The revised artifact is checked before it is considered for admission; requesting a repair does not itself establish that feasibility has been restored.

\paragraph{Challenger review.}
The challenger operates in a separate, read-only context. It examines the task specification and candidate artifact for requirements not adequately covered by the observers. A counterexample report identifies the relevant rule, affected records, observed values, and applicable bounds. The report informs subsequent refinement and is not treated as an external benchmark score.

\paragraph{Objective improvement.}
The performance prompt continues from the current files and retained incumbent. It incorporates challenger findings and requests objective improvement subject to recomputing constraints on the exported artifact. The admission check evaluates structural validity and non-regression under supported constraint checks. The request for a better objective is separate from this feasibility-based acceptance criterion, as specified in Section~\ref{sec:update}.

The industrial profiles add task-specific instructions to this phase. Datacenter search uses the two visible days and examines daily peaks and remaining resource capacity. Grid search retains the planning decisions as optimization variables and requires source replay under the 900-second solver and four-thread contract. Neither addition introduces another controller stage.

\subsection{Runtime Configuration}
\label{app:hyperparameters}
\label{app:harness_instantiation}

\begin{table}[!ht]
\caption{VERAS configuration for the 107-task ORAgentBench evaluation.}
\label{tab:app-benchmark-hyperparameters}
\centering
\small
\setlength{\tabcolsep}{6pt}
\renewcommand{\arraystretch}{1.10}
\begin{tabular}{@{}lr@{}}
\toprule
Parameter & Value \\
\midrule
Model / reasoning effort & GPT-5.4 / high \\
Concurrent task workers & 2 \\
Task-stage wall-clock ceiling & 2,700 s \\
Timeout scale & 1.00 \\
Reserved recovery fraction & 0.30 \\
Challenger fraction of remaining budget & 0.22 \\
Minimum recovery allocation & 120 s \\
Harness-side local solver cap & 180 s \\
Maximum transient transport attempts & 4 \\
Candidate SCIP call ceiling & 300 s \\
Target SCIP relative gap & $5\times10^{-4}$ \\
\bottomrule
\end{tabular}
\end{table}

The six-task ablation uses GPT-5.4/high, the same 2,700-second task-stage ceiling, public inputs, evaluator, and solver environment, with one worker per configuration. The VERAS reference is retained from the original evaluation. Runs interrupted by provider limits or transport failures were selectively rerun; completed infeasible or low-quality outputs were retained. Equal configured limits do not imply identical token consumption or elapsed time.

\begin{table*}[t]
\caption{Configurations of the selected VERAS application runs. Source replay includes setup and artifact generation; its elapsed time is not the solver-only time.}
\label{tab:app-real-hyperparameters}
\centering
\small
\setlength{\tabcolsep}{6pt}
\renewcommand{\arraystretch}{1.10}
\begin{tabularx}{\textwidth}{@{}lXX@{}}
\toprule
Parameter & Inter-datacenter placement & 54-bus network planning \\
\midrule
Model / reasoning effort & GPT-5.6 Sol / xhigh & GPT-5.6 Sol / xhigh \\
Agent wall-time limit & 19,800 s & 15,360 s \\
Timeout scale & 3.0 & 2.0 \\
Recovery reserve / challenger cap & 0.40 / 180 s & 0.40 / 480 s \\
Candidate execution contract & 120-s output generator & 900-s solver; at most four threads \\
Selected source replay & 45.842 s & 829.552 s (750-s solver limit) \\
Agent-visible instance & August 8--9, 2025 transfers & Fixed 54-bus instance \\
External evaluation & August 10, 2025 replay & Constraint and objective audit \\
\bottomrule
\end{tabularx}
\end{table*}

Both industrial profiles use the verification, routing, and admission mechanisms described in Section~\ref{sec:verification}--\ref{sec:update}. Relative to the benchmark profile, they differ in model, time limits, recovery allocation, challenger limits, and task-specific performance instructions, as listed in Tables~\ref{tab:app-benchmark-hyperparameters} and~\ref{tab:app-real-hyperparameters}. The selected grid solver leaves all eight integer decision families unfixed and uses starts derived from the task data. The agent search budget, solver-call limit, and end-to-end source-replay time measure different parts of the experiment.

\subsection{Solution Examples and Replay Results}
\label{app:cases}

\subsubsection{Scenario-Dependent eVTOL Recovery}
\label{app:evtol-case}

\begin{figure}[!ht]
\centering
\includegraphics[width=\linewidth]{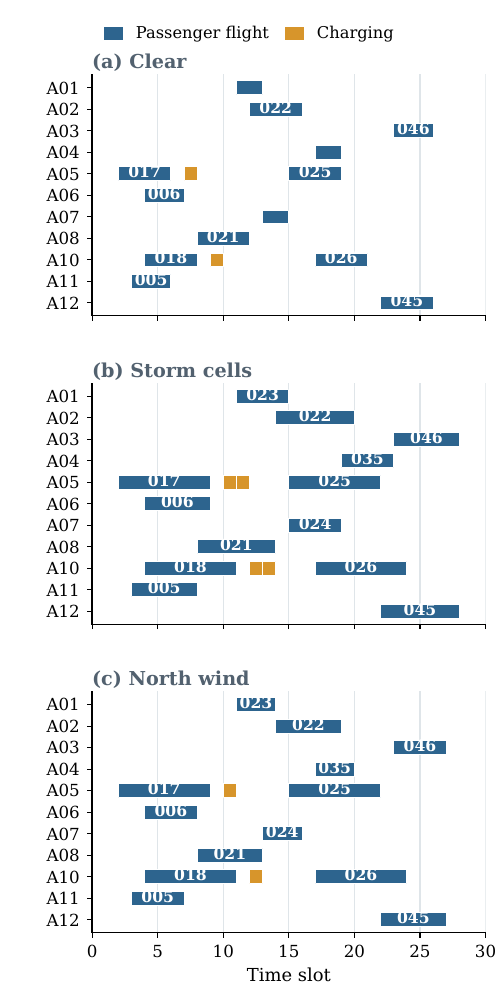}
\caption{VERAS schedules for eVTOL vertiport recovery. Blue intervals denote passenger flights and gold intervals denote charging. Flight labels are request identifiers with the leading ``R'' omitted. The accepted request set is shared across scenarios; operational schedules vary with the weather scenario.}
\label{fig:app-evtol}
\end{figure}

Figure~\ref{fig:app-evtol} shows a VERAS solution for a Hard ORAgentBench task. The artifact contains 49 first-stage acceptance decisions, of which 13 accept a request. Each accepted request has one flight in each of three weather scenarios, giving 39 flight rows; the artifact also includes eight charging rows. The acceptance decisions are shared across scenarios, whereas flight and charging schedules can vary. The timeline observer examines 33 aircraft--scenario timelines and reports no violations of release times, turnaround requirements, non-overlap, or location continuity within its implemented scope. This observation does not certify constraint families outside that scope.

\subsubsection{Inter-Datacenter Placement}

\begin{figure}[!ht]
\centering
\includegraphics[width=\linewidth]{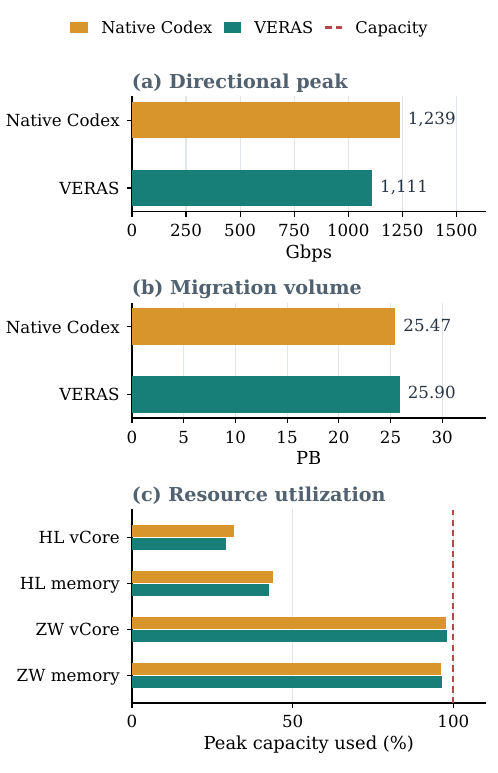}
\caption{August 10 replay of archived VERAS and the later native Codex run: (a) maximum directional bandwidth peak; (b) migration volume; and (c) resource utilization. All four utilization ratios are below the capacity limit.}
\label{fig:app-datacenter}
\end{figure}

The archived VERAS assignment in Table~\ref{tab:bandwidth} has August 10 directional peaks of 1,110.603 Gbps from Huailai to Zhongwei and 907.866 Gbps in the reverse direction. Peak vCore and memory utilization ratios are 0.294 and 0.428 in Huailai, and 0.979 and 0.965 in Zhongwei. Figure~\ref{fig:app-datacenter} compares these outcomes with the later native Codex run. In the recorded source replays, both programs produce resource-feasible assignments and regenerate their Parquet artifacts byte-for-byte from the agent-visible inputs. Migration is reported separately from the bandwidth objective.

\paragraph{Additional VERAS run.}\label{app:dc-replicate}
A new VERAS run with the same performance instructions yields 1,567.303 Gbps on August 10, compared with 1,238.625 Gbps for native Codex. Both VERAS runs and the native run satisfy the resource constraints and regenerate their assignments within 120 seconds. This additional run reverses the ranking of the archived comparison; the reported 10.34\% reduction is therefore specific to the selected archived run.

\subsubsection{54-Bus Planning and Fault Operation}
\label{app:grid-case}

Table~\ref{tab:adn} reports a separately selected VERAS run with annualized cost USD~39.424M. The VERAS arm executed alongside native Codex instead obtains USD~42.613M. The selected VERAS and USD~45.690M native solutions use the same task instance, LLM backend, evaluator, four-thread limit, and 900-second solver contract; they are not outputs of the same paired comparison. With a 750-second solver limit for source replay, their end-to-end times are 829.552 and 823.924 seconds, respectively. Both replayed artifacts pass the independent constraint and objective audit.

\begin{figure}[!ht]
\centering
\includegraphics[width=\linewidth]{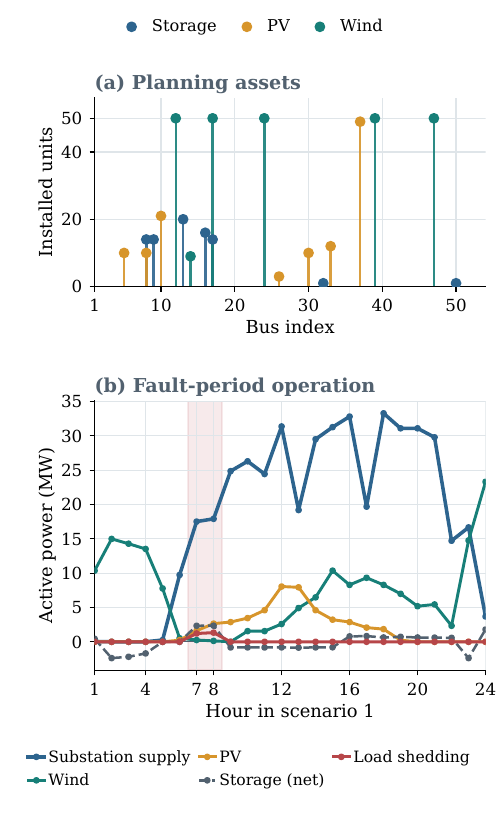}
\caption{Selected VERAS grid solution: (a) installed integer units by technology and bus; and (b) scenario-1 active power, with the fault interval (hours 7--8) shaded. Positive net storage power denotes discharge. Load shedding is reconstructed from the saved decisions.}
\label{fig:app-power-grid}
\end{figure}

The selected VERAS solution installs storage at buses $\{8,9,13,16,17,32,50\}$, photovoltaics at $\{5,8,10,26,30,33,37\}$, and wind at $\{12,14,17,24,39,47\}$. Figure~\ref{fig:app-power-grid} shows the installed integer units and aggregate active power in scenario 1, converted to MW using a 10-MVA base. Independently reconstructed load shedding during fault hours 7 and 8 is 1.209 and 1.330 MW, respectively.

The selected run reports a best bound of USD~37.404M and a 5.122\% solver gap for its optimization model. This is a feasible incumbent, not a proof of global optimality. The objective reconstructed from the saved decisions agrees with the reported value; the integrality residual is zero and the maximum normalized constraint residual is $8.54\times10^{-10}$. The recorded source replay reproduces the NPZ artifact byte-for-byte. These checks validate the fixed program and exported solution, not the repeatability of the preceding agent search.

\subsection{Released Materials and Reproducibility}
\label{app:reproducibility}

The benchmark release contains task-level scores for all 107 instances and raw event streams for 34 tasks. The application records include solver source code, task-input hashes, and replay audits. The manuscript project also includes aggregate datacenter and grid summaries used by the case-study plotting script. These summaries support checks of the reported aggregate values, but do not replace the original decision artifacts or constraint audits.

Operational input rows and generated Parquet/NPZ artifacts are not redistributed. Consequently, independently reproducing the industrial results requires access to the original task data in addition to the released programs. The replay results above describe checks performed with those inputs available; they do not imply that the public archive alone is sufficient to rerun the applications.

\end{document}